# Free-form conformal metasurfaces robustly generating topological skyrmions

Yang Fu, Rensheng Xie, Nilo Mata-Cervera, Xi Xie, Ren Wang, Xiaofeng Zhou, Helin Yang * and Yijie Shen *

Yang Fu, Xiaofeng Zhou, Helin Yang

College of Physical Science and Technology, Central China Normal University,

Wuhan 430079, China

E-mail: *emyang@ccnu.edu.cn

Rensheng Xie, Nilo Mata Cervera, Xi Xie, Yijie Shen

Centre for Disruptive Photonic Technologies, School of Physical and Mathematical Sciences, Nanyang Technological University, Singapore 637371, Republic of Singapore

E-mail: *yijie.shen@ntu.edu.sg

Ren Wang

Institute of Applied Physics, University of Electronic Science and Technology of China, Chengdu 611731, China.

Yangtze Delta Region Institute (Huzhou), University of Electronic Science and Technology of China, Huzhou 313098, China.

Yijie Shen

School of Electrical and Electronic Engineering, Nanyang Technological University, Singapore 639798, Republic of Singapore.

Rensheng Xie

Department of Broadband Communication, Peng Cheng Laboratory, Shenzhen, 518000, China

Skyrmions are topologically stable vector textures as potential information carriers for high-density data storage and communications, especially boosted by the recently emerging meta-generators of skyrmions in electromagnetic fields. However, these implementations always rely on planar, rigid designs with stringent fabrication requirements. Here, we propose the free-form conformal metasurface generating skyrmions towards future wearable and flexible devises for topological resilience light fields. Furthermore, we experimentally tested the outstanding topological robustness of the skyrmion number under different disorder degrees on the metasurface. This work promotes the development of flexible compact skyrmion-based communication devices and demonstrates their potential to improve the quality of space information transmission.

# 1. Introduction

Skyrmions[1–4] are topologically protected quasiparticles originally proposed in nuclear and condensed matter physics. Owing to their rich internal structure, skyrmions[5–8] have been extensively explored in various condensed matter physics[9], including quantum Hall magnets[10], Bose-Einstein condensation[11,12], or liquid crystals[13,14]. More recently, the concept has been extended to optics, where optical skyrmions [15–17] are realized as structured vector fields of light, providing a direct analogue to magnetic skyrmions. Through precise control over amplitude, phase, and polarization, a wide variety of optical skyrmions have been demonstrated, including those associated with Stokes vectors[18,19], spin angular momentum[20], momentum[21] and electromagnetic field skyrmions[22]. These structured excitations possess distinctive properties [23,24]—such as resilience to perturbations [25–27], self-healing capability [28,29]—that make them particularly suitable for information transfer in free space [30]. Importantly, these features also suggest new opportunities for multiplexing, high-capacity communication, and topologically protected photonic systems.

The ability to harness skyrmions for communication and photonic technologies requires practical and versatile generation devices. In contrast to the bulky SLM, meta-generators emerged for practical applications, such as metasurfaces[31–34], metafibers [35,36], and bound states in the continuum (BIC) slabs [37,38]. However, most reported generators depend on auxiliary free-space optics or are fabricated on rigid, planar architectures and therefore offer limited conformability. More critically, existing generators often lack structural robustness and interference tolerance. Such limitations significantly hinder their scalability and integration into emerging technologies, particularly in the context of wearable photonics, conformal devices, and multi-channel wireless communication. Accordingly, there is a pressing need for flexible, interference-tolerant generators that can leverage the topological stability of skyrmions to efficiently transmit information in free space.

Here, a versatile method is proposed to address the aforementioned limitations. Taking Stokes skyrmions as an example, we propose a flexible metasurface [39–41] platform capable of robustly generating free-space skyrmions. Unlike previous rigid designs, our approach allows skyrmions to be generated on conformal structures, such as planes, cylinders, and triangles, without compromising the skyrmions number. Furthermore, even after randomly removing over 60% of the metasurface units, the resultant skyrmion number exhibits exceptional robustness under the presence of defects, demonstrating that the proposed system maintains functionality even under substantial structural perturbations. This not only paves the way for future developments in topological photonics and wearable microwave devices, but also exhibits great potential as robust, miniaturized information carriers for next-generation wireless networks.

# 2. Concept and Theory

## 2.1. Characterization of Skyrmions

**Figure 1** shows a perfect Néel-type skyrmion with the intensity distributions of its circularly polarized components. Stokes skyrmions in the microwave band are generated using spin-decoupled metasurfaces with both planar and conformal geometries. The metasurface is designed to independently tailor optical vortex beams and then focused them through two spin-decoupled channels. This enables the superposition of Laguerre–Gaussian modes with different topological charges and polarizations. In Reference[42], the authors propose a practical scheme for the generation of tunable optical skyrmions in free space by structuring the Stokes vector of a vector beam. The Stokes vector $\mathbf{s} = (s_1, s_2, s_3)$ can represent any polarization state as a point on the surface of a sphere of unit radius, known as the Poincaré sphere. In the spherical coordinate system, the light field can be mathematically expressed as: $\mathbf{\psi} = \cos(\beta/2)e^{-i\alpha/2}\mathbf{R} + \sin(\beta/2)e^{i\alpha/2}\mathbf{L}$, where $\mathbf{R}$ and $\mathbf{L}$ represent the right-handed circularly polarized (RCP) and left-handed circularly polarized (LCP) waves, $\alpha$ and $\beta$ are the spherical and azimuthal angle in the sphere. The corresponding Stokes vector components are defined as $\mathbf{s} = (\cos\alpha\sin\beta, \sin\alpha\sin\beta, \cos\beta)$. To shape a skyrmion texture using the Stokes vectors, we need to independently modulate two spatial modes in two orthogonal polarization states, so that the polarization states varies over the transverse plane. The resulting vector beam typically takes the form:

$$\mathbf{\psi}(x,y) = \psi_R(x,y)\mathbf{R} + \psi_L(x,y)\mathbf{L} \qquad (1)$$

Here, $\psi_R(x,y)$ and $\psi_L(x,y)$, are the scalar wavefunctions of RCP and LCP polarization components. At its simplest we can choose these components to be a fundamental Gaussian mode and a Laguerre-Gaussian mode with topological charge $l=1$, that is $\psi_R(x,y) = LG_{00}$ and $\psi_L(x,y) = LG_{01}$, with their intensity profiles depicted at Figure 1(a) and (b) respectively. The resulting Stokes vector field corresponds to a Néel-type skyrmion, a stereographic projection of the Poincaré sphere into the transverse plane. As it has been discussed in References[37], Laguerre-Gaussian beams are a more flexible basis than Bessel beams given their simplicity in their experimental realization.

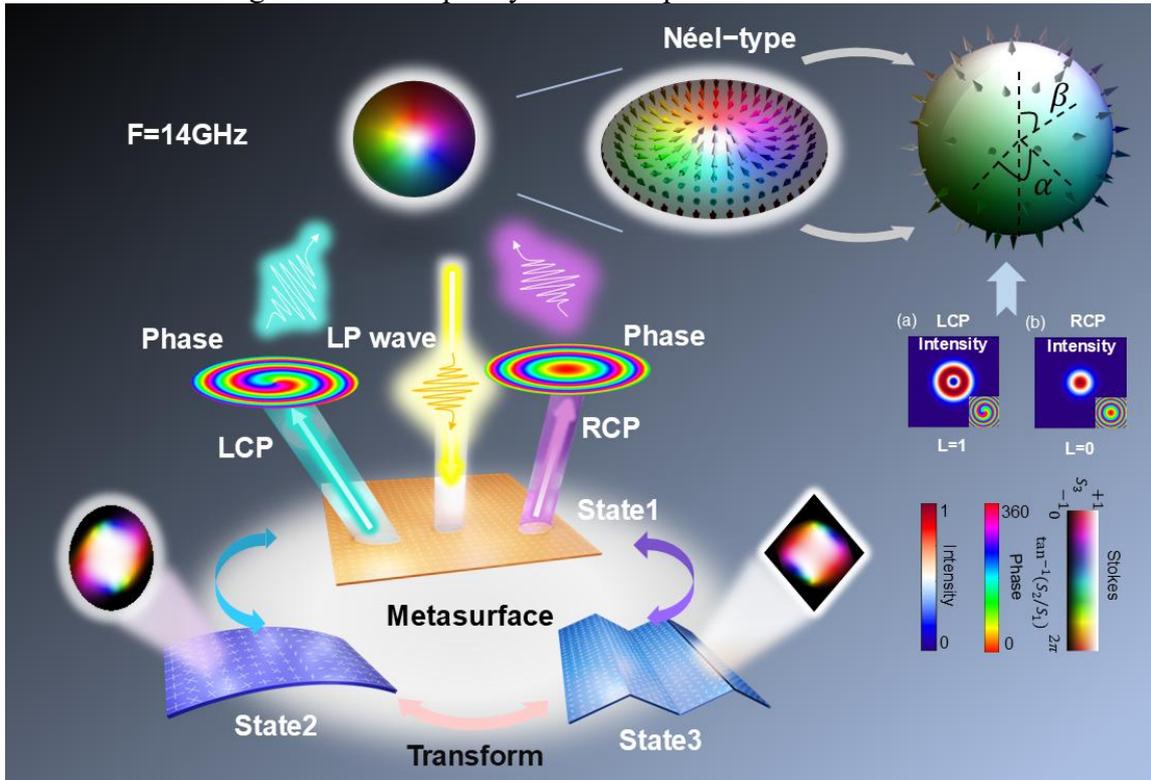

**Figure 1.** Conformal metasurface for generating microwave band skyrmion. When the free-form metasurface is converted from a planar structure (state 1) to a cylindrical (state 2) or triangular one (state 3), the skyrmion texture changes from a circle to an ellipse or a rhombus under the incidence of a linearly polarized electromagnetic wave. All these structures exhibit strong robustness for generating Stokes skyrmions. (a) shows the electric field intensity of LCP wave, (b) shows the electric field intensity of RCP wave, and (c) shows the three-dimensional Stokes vector in the surface of the Poincaré sphere, and its stereographic projection onto the transverse plane.

The topological properties of a skyrmion beam are characterized by the skyrmion number ($N_{sk}$) [42]:

$$N_{sk} = \frac{1}{4\pi} \iint_\sigma \boldsymbol{n} \cdot (\frac{\partial \boldsymbol{n}}{\partial x} \times \frac{\partial \boldsymbol{n}}{\partial y}) dxdy \tag{2}$$

where $\boldsymbol{n}$ is the normalized Stokes vector and $\sigma$ is the region confining the skyrmion. When expressed in terms of the spherical angles in the unit Poincaré sphere, the Stokes vector is given by $\boldsymbol{n} = (\cos\alpha(\phi)\sin\beta(r), \sin\alpha(\phi)\sin\beta(r), \cos\beta(r))$, where $\phi$ and $r$ the azimuthal and radial coordinates in the transverse plane. Equation (2) can be solved and the skyrmion number simply reads $N_{sk} = -\frac{1}{4\pi} \cos\beta(r)|_{r=0}^{r=R_s} \cdot \alpha(\phi)|_{\phi=0}^{\phi=2\pi}$.

This topological number quantifies the total solid angle of the Poincaré sphere that is spanned by the vector field in the transverse plane. When the skyrmion number is 1, the entire surface of the Poincaré sphere is covered exactly once, in other words, all transverse polarization states are present at the transverse plane. This property significantly enhances the channel capacity of optical and microwave systems and underscores the potential of skyrmions in multi-channel communication.

## 2.2. Planar metasurfaces generating skyrmions

The metasurface structure consists of five layers, as shown in **Figure 2**(a) and (b). The first and second layers consist of a 0.035 mm-thick copper layer and a 0.2 mm-thick F4B (polytetrafluoroethylene) dielectric layer ($\varepsilon_r$ = 3.0, and $\tan d$ = 0.001). The use of ultrathin F4B enables mechanical flexibility, allowing the planar metasurface to be bend, which is essential for investigating the quality of skyrmions on conformal surfaces. The third layer is a 2.1 mm resin material ($\varepsilon_r$ = 3.0, and $\tan d$ = 0.0375). To discuss the properties of Stokes skyrmions on flexible metasurfaces, we use 3D printing technology to make conformal media with arbitrary curvature, for which we use a resin structure in the simulation. The fourth and fifth layers are a F4B layer with a thickness of 0.2 mm and a copper layer with a thickness of 0.035 mm, respectively.

According to spin-decoupled theory (Supporting Information, note1), the propagation phases along the X- and Y-directions can be controlled by varying the lengths $L_x$ and $L_y$ of the cross structure, as shown in Figure 2(c). The rotation angle $\theta$ can be used to control the geometric phase. We performed a scan optimization of the unit structure parameters, obtaining the appropriate sizes in the unit cell which allow for $2\pi$ phases coverage. Figure 2(c) shows the size and phase of the eight different meta-atoms, quantized using 2-bit coding.

The electric field amplitude and phase distributions of the planar spin-decoupled metasurface were obtained using full-wave simulations performed in CST Microwave Studio. Details of the simulation setup and phase modulation strategy are provided in Supporting Information (note 1). Figure 2(d-f) illustrates the results under the incidence of a linearly polarized wave generating Stokes skyrmions with target skyrmion numbers $N_{sk}$ = 1,2 and 3, respectively. The RCP component does not carry any vortex phase and it is focused into a central spot, while the LCP component, shows the typical doughnut-like intensity distribution. The corresponding skyrmion numbers calculated via Equation (2) are $N_{sk}$ = 0.9531, 1.9893 and 2.9096, aligning very closely with the target values. The Stokes vector maps (Figure 2(d4–f4)) show the continuous mapping from the Poincaré sphere to the transverse plane. These results verify that the metasurface can successfully generate Stokes skyrmions with skyrmion numbers from 1 to 3 with remarkable quality. Supporting Information (note 1) discusses the case of planar metasurfaces generating higher-order Stokes skyrmions.

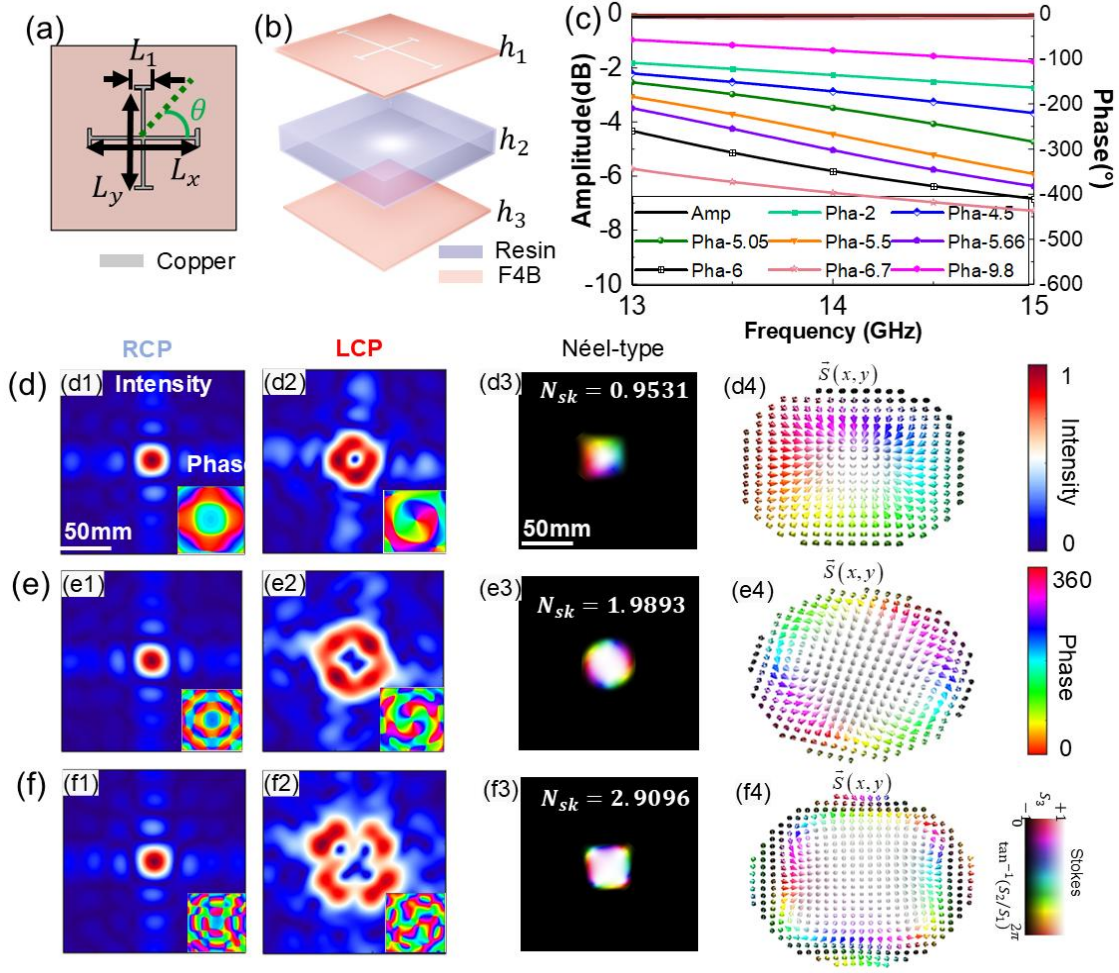

**Figure 2.** (a) Upper layer of the metasurface unit, (b) Overall view of the metasurface unit, (c) Plot of amplitude and phase of metasurface units. The calculated amplitude, phase, and texture diagrams, as well as the electric field vector diagrams, are shown for topological numbers 1, 2, and 3 skyrmions under a planar structure, as shown in Figure 2(d), (e) and (f). Scalebar:50 mm.

## 2.3. Conformal metasurfaces robustly generating skyrmions versus distortion

To further investigate the quality of the generated Stokes skyrmions using non-planar structures, we extend our design to conformal metasurfaces with cylindrical and triangular geometry. These flexible structures introduce spatially varying path lengths, which must be accurately compensated to preserve the intended polarization and phase distributions.

For the cylindrical configuration, shown schematically in **Figure 3**(a), phase compensation is implemented based on generalized Snell's law (as described in Equation (S2)). This approach enables precise control over the reflected wavefront, ensuring the correct generation of Stokes skyrmions despite the distortion induced by the surface curvature. $f(x)$ represents the cross-section function of the cylinder, $g(x)$ represents the cross-section function of the metasurface, and $h$ represents the height of the metasurface which is used in the calculation of the optical length. In Figure 3(b), P represents the distance from the center of the skyrmion to the center of the metasurface. When the size or curvature of the metasurface changes, the function $f(x)$ changes as well. $R$ is the radius of curvature after bending, and the bending angle of the metasurface is defined as $b$, which is 80° in this case. Here, the functions $f(x)$ and $g(x)$ used in this paper are defined as such: $f(x)=\sqrt{200^2-x^2}$, $g(x)=\sqrt{200^2-x^2}-k0$, where $k0$ represents a const that is related to the selection of the reference plane.

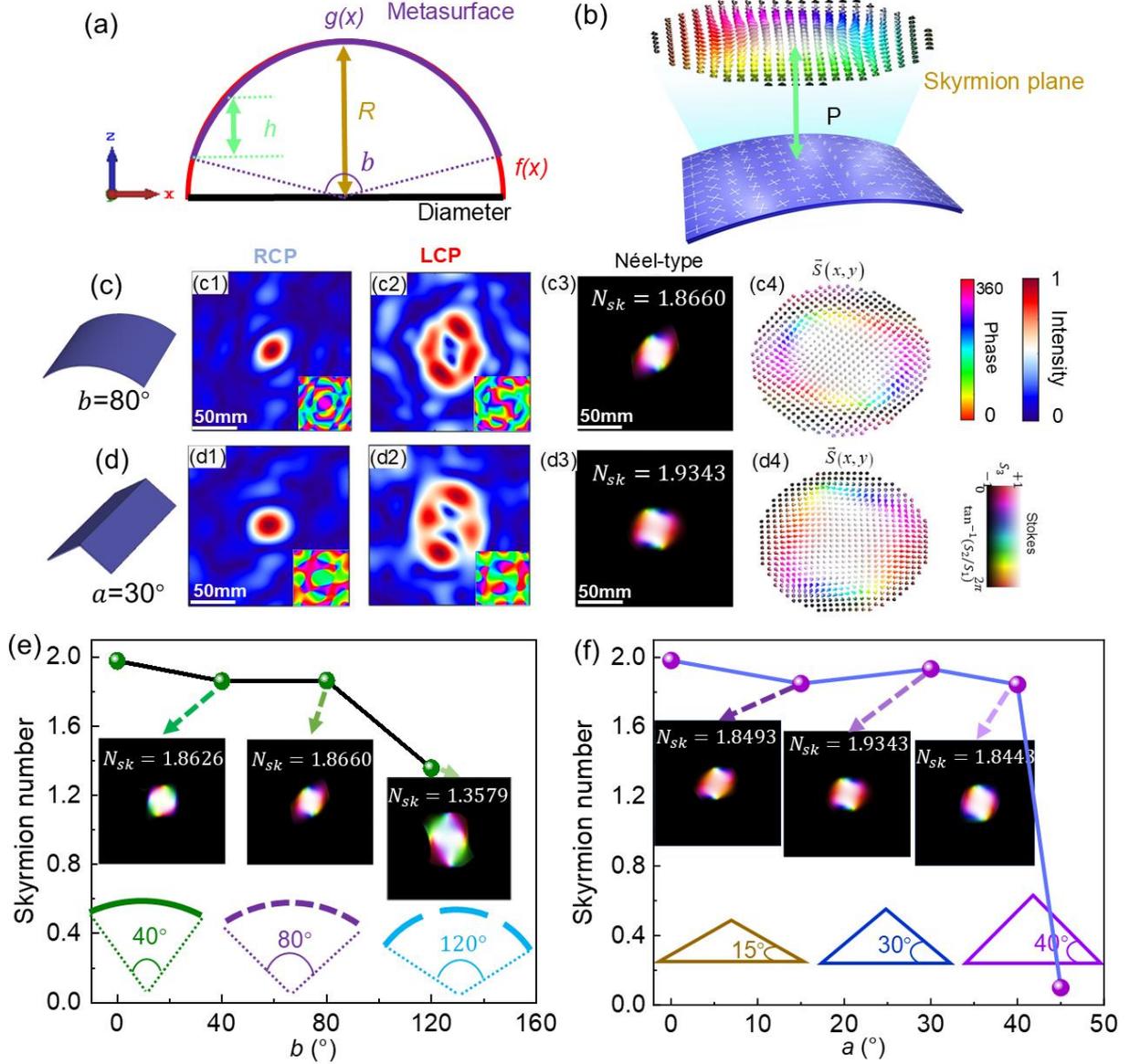

**Figure 3.** (a) Schematic diagram of the metasurface bending into a curved surface. (b) Diagram of the skyrmion generation. Amplitude, phase, and Stokes vector texture diagrams for the $N_{sk}$= 2 skyrmions are shown in (c) for $b$ = 80°, and in (d) for triangular structures with $a$=30°. (e) and (f) show the skyrmion numbers and corresponding Stokes textures generated by conformal metasurfaces with cylindrical and triangular geometry for varying geometric parameters $a$ and $b$. Scalebar:50 mm.

The generation of Stokes skyrmions from triangular metasurfaces similarly requires precise calculation of optical length differences, as in the case of cylindrical structures. Detailed calculations are given in the Supporting Information (note 2). Figure 3 compares the skyrmion textures produced by cylindrical and triangular conformal metasurfaces. The skyrmion number obtained from the cylindrical configuration is 1.8660 for $b$=80°, while the triangular metasurface yields a skyrmion number of approximately 1.9343 for $a$=30°. Further analysis on the the skyrmion number as a function of the bending angle and geometry is presented in the Supporting Information (note 3).

Among other structural parameters, the radius of curvature has an important impact on the quality of the generated skyrmions. In practical applications, the radius and bending angle can be tuned to accommodate different conformal designs. The corresponding phase compensation values can also be accurately calculated using Equations (S1) and (S2). As shown in Figure 3(e) and (f), when the size of the metasurface array is fixed, stronger bending corresponds to larger *b* angles. We simulated the response of the structure and calculated the skyrmion numbers of the field for different angles. In the cylindrical structure, when the bending angle exceeds 90°, the incidence angle of the units at the edge of the metasurface array exceeds 45°, which results in a stronger phase distortion and a drop in the skyrmion number. Similarly for the triangular structure, the generated skyrmions can maintain a good quality when the angle *a* does not exceed 45°. The disruption of the skyrmion quality at higher bending angles is related to the strong phase distortion, which severely degrades the quality of the vortex beam when *b* exceeds 90° or *a* exceeds 45°. For smaller angles, the generated skyrmions number remain remarkable stable.

Both cylindrical and triangular structures effectively generate skyrmion textures, whose degree of coverage of the Poincaré sphere closely matches the ideal values. Although the electric field of skyrmions changes on conformal metasurfaces, their topological number remains stable. These results indicate that the influence of surface geometry on skyrmion number is relatively limited, provided that proper phase compensation is implemented. Therefore, skyrmions can be reliably generated on metasurfaces of arbitrary curvature, offering great flexibility for practical implementation in multi-channel microwave communication systems.

## 3. Experimental Results

### 3.1 Generation and detection of skyrmions from conformal metasurfaces

To validate the properties of skyrmions on conformal geometries, curved metasurfaces were fabricated and tested. The top and bottom metasurfaces were fabricated using conformal materials, and the middle layer was fabricated using 3D printing, as shown in **Figure 4**. The use of 3D printing allows for the realization of arbitrary metasurface shapes, enabling a fully conformal integrated design adaptable to diverse geometric configurations. During the measurements process, due to the design of the reflective metasurface and the fact that the test probe is located on the wall with limited movement, there will be an occlusion of the incident electromagnetic wave with the horn. To minimize the effect of the setup configuration on the generated skyrmions the incident wave was set at a small angle. The front view of the experiment is shown in Figure 4(a), where the transmission and reception of electromagnetic waves are performed by two horn antennas. The relative positioning of the antennas with respect to the test plane is shown in Figure 4(b). For more details regarding the experimental setup please refer to the Supporting Information (note 4).

Figure 4 (c) and (d) shows the experimental results obtained from the planar and conformal metasurfaces, respectively. We measured the near-field distribution after reflecting on the metasurface for both LCP and RCP components. The LCP component presents asymmetry in the intensity distribution and a splitting of the second order phase singularity into two phase singularities, which is expected from the imperfections in the fabrication and the input illumination. Moreover, the polarization profiles present also visible deformations with respect to the ideal cylindrically symmetric case. Nevertheless, although the patterns are not symmetric, the experimentally retrieved skyrmion numbers reached 1.8328 and 1.8098, which is in close agreement with the ideal results. These results show the remarkable robustness of the skyrmion number even in the case of strong distortion of the polarization field, which may have promising applications for communications in the microwave band.

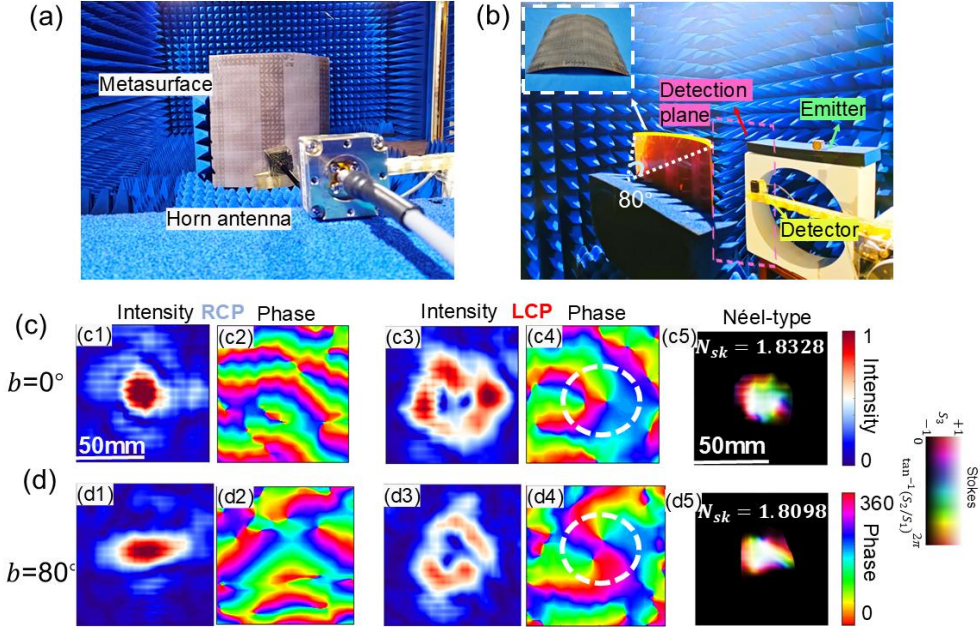

**Figure 4.** Metasurface experimental setup (a) front view and (b) back view. (c) Measurement results for the planar metasurface with $b=0$, including RCP intensity (c1), and phase (c2), LCP intensity (c3), and phase (c4), and the reconstructed Stokes skyrmion texture with skyrmion number $N_{sk}$=1.8328 (c5). (d) Measured results for the conformal metasurface with $b=80$, including RCP intensity (d1), and phase (d2), LCP intensity (d3), and phase (d4), and skyrmion texture with $N_{sk}$=1.8098 (d5). The dashed circles in (c4) and (d4) denote the spatial region containing the phase vortices of topological charge $l$=2. Scalebar: 50 mm.

## 3.2 Exploration of Topological Robustness

In both simulation and experimental results presented above, the metasurfaces—whether planar or conformal—successfully generated skyrmions with high topological numbers, demonstrating excellent agreement between theory and experiments. However, for real-world applications such as wireless communications, sensing, or field encoding, the robustness of skyrmion generation under structural imperfections is of paramount importance. Skyrmions are known for their intrinsic topological robustness, allowing them to maintain distinguishable topological features even when the underlying field distribution is locally perturbed.

To examine this property in the context of metasurface-based generation, we investigated the evolution of the skyrmion number when some metasurface unit cells were removed along the central region and randomly removed, as shown in **Figure 5**(a). Taking the planar metasurface as a representative example, we carried out both theoretical calculations and full-wave simulations to analyze the skyrmion number and texture under different unit-cell removal ratios. To ensure statistical accuracy, the theoretical skyrmion number was obtained by averaging over 100 independent random configurations for each removal percentage. Calculations were performed using a 10% sampling interval. The figure shows a distribution diagram of partial percentages: 0%, 20%, 40%, 60%, 80% and 100%. Both numerical simulations and microwave experiments were carried out to quantify the degree to which the topological charge is preserved under such perturbations. The corresponding metasurface structures for each removal percentage are provided in Supporting Information (note 4).

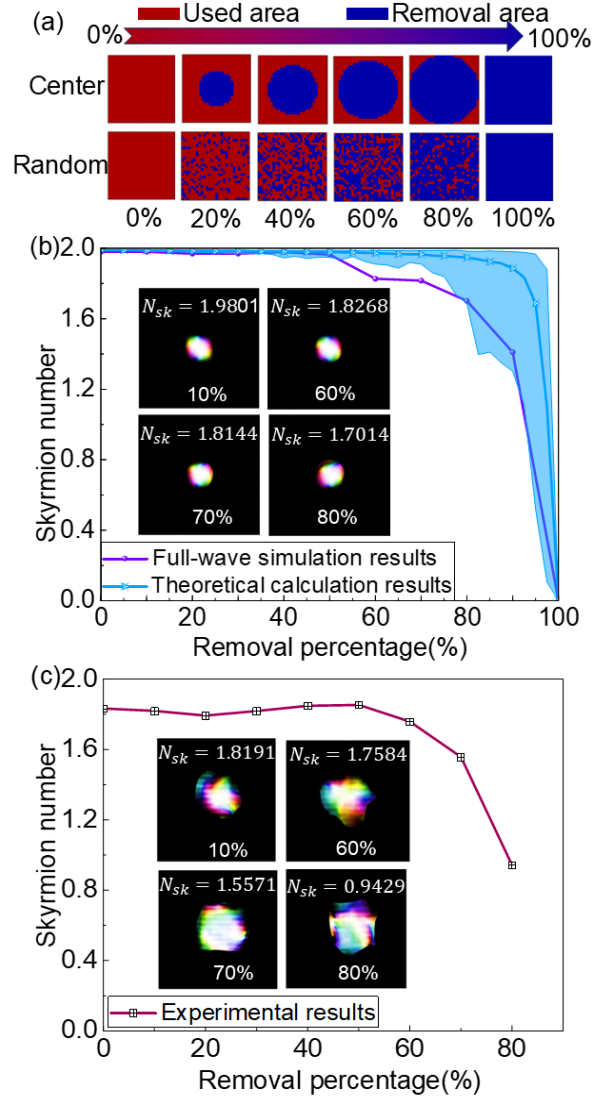

**Figure 5.** (a) Metasurface units are removed along the central region (first row) or randomly (second row). The removal ratio increases from left to right, respectively, at 0%, 20%, 40%, 60%, 80%, and 100%. Comparison of (b) simulated and (c) experimental results of the skyrmions number obtained after randomly removing a certain percentage of the units from the planar metasurface. The blue region in (b) showcases the standard deviation of the skyrmion number after 100 stochastic realizations of random removal of the metasurface unit cells.

Figure 5(b) shows a direct comparison between the theoretically calculated results (blue curve) and full-wave simulation results (purple curve). The theoretical results assume that the unit cells are removed, yielding zero amplitude of the reflected wave at that pixel, while the response of the remaining unit cells is the same as in the ideal case with no unit cells removed. The removal ratio ranges from 0% to 100%. The blue curve represents the average skyrmion number over 100 randomized metasurface realizations at each removal ratio. The results indicate that the skyrmion number can remain above 1.8 even when the percentage of units removed reached 70%, indicating remarkable topological resilience.

The full-wave simulation results closely match the theoretical predictions for removal rates below 45%. Beyond 50% deficiency, discrepancies emerge due to significant energy leakage, since most incident microwaves pass through the metasurface without being reflected, yielding zero amplitude in the reflected wave at that unit cell which impacts the response of neighboring unit cells significantly affecting the stokes vector distribution in the near field. Consequently, the skyrmion number obtained from simulation is lower than that from theoretical calculations. Additional deviations can be attributed to propagation losses and spatial quantization errors in the simulation grid. Nonetheless, the simulated curve remains consistent with the overall trend of the theoretical data.

Figure 5(c) presents the experimental measurements of skyrmion number under varying removal ratios. The corresponding fabricated metasurfaces with different defect levels are also included in Supporting Information (note 4). The skyrmion number stays stable at around 1.8 for removal percentages below 50%, consistent with both theoretical and simulated results. As the removal percentage increases to 60%, the topological number begins to decline significantly, reaching a minimum below 1.0 at 80%, indicating the significant degradation of the topological features.

These results not only confirm the strong topological robustness of skyrmions in the microwave band but also constitute the first experimental verification of such stability at these frequencies. Full experimental data are provided in Supporting Information (note 4). This finding highlights the great potential of microwave skyrmions for robust field encoding, multi-channel wireless communication, and energy-transfer applications, even under severe structural degradation.

## 4. Conclusion

In this work, we theoretically proposed and experimentally demonstrated a method for generating Stokes skyrmions in the microwave band using free-form spin-decoupled metasurfaces. By shaping two scalar modes with different topological charges in orthogonal circular polarizations, we successfully generated Stokes skyrmions with controllable topological numbers. Furthermore, the topological nature of the generated skyrmions were verified both through simulations and by experimental measurements in a microwave anechoic chamber. The robustness of the retrieved skyrmion number under random structural deletions was evaluated, revealing that even with over 60% of the metasurface units removed, the generated skyrmion can still maintain a skyrmion number close to the theoretical value.

This work introduces a new degree of freedom for generating skyrmions in the free-form structures and opens promising opportunities for applications in flexible communication systems, adaptive sensing platforms, and intelligent field recognition. In wireless communication[43], the introduction of skyrmion-based vector encoding may enable multi-channel and high-capacity information transmission, with robustness against signal degradation caused by structural imperfections. For microwave sensing[44,45], the Stokes textures of free-space skyrmions allows for enhanced spatial resolution, polarization-selective detection, and improved environmental adaptability. In pattern recognition and imaging systems[46–49], the topologically stable and flexible field distribution can support robust feature extraction under deformation or partial signal loss, which is particularly beneficial in wearable and conformal electronics.

Moreover, the demonstrated topological robustness—the ability of skyrmions to maintain their topological features despite large-scale random deletions of the metasurface structure—offers an important advantage for real-world applications. It ensures system functionality even in the presence of fabrication defects, degradation, or environmental disturbances. Besides improving device reliability, this property also enables cost-effective and fault-tolerant design of future skyrmion-based microwave photonic systems[50,51].

Overall, this research establishes a flexible, stable, and scalable route to topological field engineering in the microwave domain. Moreover, it provides a solid foundation for the integration of skyrmions into next-generation communication, sensing, and intelligent microwave systems.

## 5. Experimental Section

The experimental samples were fabricated from multilayer media. The first layer consists of a double-sided PTFE (polytetrafluoroethylene: $\varepsilon_r$ = 3.0, and tan $d$ = 0.001) copper-clad plate. Excess copper was then etched using a chemical reagent to form a patterned metasurface. The resin ($\varepsilon_r$ = 3.0, and tan $d$ = 0.0375) was synthesized from organic materials that were melted at high temperatures and shaped into the desired thickness and dimensions using 3D printing. After the structure was formed, the layers were carefully bonded. Experiments were conducted in a microwave anechoic chamber using a horn antenna for near-field measurement.

**Supporting Information**

Supporting Information is available from the Wiley Online Library or from the author.

**Acknowledgements**

This work was financially supported by self-determined research funds of CCNU from the colleges' basic research and operation of MOE (CCNU25ai042), Basic Research Funding for Central Universities (Excellent Innovation Project) under Grant 30106250136, Nanyang Assistant Professorship Start Up Grant, Singapore Ministry of Education (MOE) AcRF Tier 1 grants (RG157/23 & RT11/23), and Singapore Agency for Science, Technology and Research (A*STAR) MTC Individual Research Grants (M24N7c0080)

**Conflict of Interest**

The authors declare no conflict of interest.

**Data Availability Statement**

The data that support the findings of this study are available from the corresponding author upon reasonable request.

Received: ((will be filled in by the editorial staff))

Revised: ((will be filled in by the editorial staff))

Published online: ((will be filled in by the editorial staff))